\pdfoutput=1
\documentclass[10pt, conference, letterpaper]{IEEEtran}
\IEEEoverridecommandlockouts
\usepackage{cite}
\usepackage{amsmath,amssymb,amsfonts}
\usepackage{algorithmic}
\usepackage{graphicx}
\usepackage{textcomp}
\usepackage{xcolor}
\usepackage{svg}
\usepackage{braket}
\usepackage{booktabs}
\usepackage{caption}
\def\BibTeX{{\rm B\kern-.05em{\sc i\kern-.025em b}\kern-.08em
    T\kern-.1667em\lower.7ex\hbox{E}\kern-.125emX}}

\usepackage{graphicx}
\graphicspath{ {./fig/} }

\usepackage{multirow}
\usepackage{booktabs}

\usepackage{cite}
\usepackage{amsfonts}
\usepackage{amssymb}   
\usepackage{amsmath}
\usepackage{algorithmic}
\usepackage{adjustbox}
\usepackage{caption}
\usepackage{wrapfig}
\usepackage{enumitem}

\usepackage{xcolor}
\usepackage[ruled,vlined]{algorithm2e}

\SetCommentSty{mycommfont}
\usepackage{soul}
\usepackage{mathtools}

\usepackage{colortbl}

\usepackage{makecell}

\usepackage{textcomp}

\begin{document}
\abovedisplayskip=5pt
\abovedisplayshortskip=5pt
\belowdisplayskip=5pt
\belowdisplayshortskip=5pt

\title{HarmQ: Harmonic Backdoor Attacks Against Quantum Neural Networks}

\author{
Junrui Zhang$^{1}$, Zemin Chen$^{1}$, Chunsheng Xin$^{2}$, Hongyi Wu$^{3}$ and Rui Ning$^{1}$ \\[0.6em]
$^{1}$Department of Computer Science, Old Dominion University, Norfolk, VA, USA \\
$^{2}$Department of Computer Science, Iowa State University, Ames, IA, USA \\
$^{3}$Department of Electrical and Computer Engineering, University of Arizona, Tucson, AZ, USA
}



\maketitle

\begin{abstract}
\noindent Quantum Neural Networks (QNNs) have emerged as a promising paradigm for quantum machine learning in the Noisy Intermediate-Scale Quantum (NISQ) era, leveraging quantum phenomena such as superposition and entanglement to process information in exponentially large Hilbert spaces. However, QNNs inherit critical security vulnerabilities from classical neural networks, particularly susceptibility to backdoor attacks. Existing attack methods designed for classical systems fail against QNNs due to quantum-specific constraints: aggressive downsampling required by limited qubit resources destroys conventional triggers, while the spectral learning bias of parameterized quantum circuits (PQCs) restricts learnable patterns. To tackle this, we present \textbf{HarmQ}, a quantum-native backdoor attack that exploits PQCs' inherent Fourier decomposition bias through harmonic trigger patterns. Our approach employs sinusoidal perturbations on coarse grids with block-uniform structure, ensuring survival through downsampling while aligning with PQCs' preference for low-frequency components. This enables effective backdoor injection under realistic black-box conditions where attackers access only training data. Experiments on MNIST and Fashion-MNIST demonstrate that HarmQ achieves attack success rates exceeding 99\% while maintaining over 90\% clean accuracy, significantly outperforming existing methods including BadNets (2.77\% ASR), Watermark (7.96\% ASR), Q-FGSM (44.32\% ASR) and QUAP (3.40\% ASR). Parametric t-SNE visualizations of quantum state representations confirm that harmonic triggers create distinctly separated clusters, evidencing HarmQ as a fundamental security threat for QNNs.

\end{abstract}


\vspace{-2mm}\section{Introduction}\vspace{-2mm}

Machine learning has become a cornerstone of modern intelligent systems, achieving remarkable progress in computer vision, natural language processing, and autonomous systems. Recently, quantum computing has emerged as a transformative paradigm that could fundamentally enhance machine learning capabilities. By leveraging quantum phenomena such as superposition and entanglement, quantum systems can embed classical data into exponentially large Hilbert spaces, creating feature representations that capture complex patterns beyond the reach of classical methods~\cite{havlivcek2019supervised}.



In the current noisy intermediate-scale quantum (NISQ) era~\cite{preskill2018quantum}, where quantum devices are limited by noise and shallow circuit depths, variational quantum algorithms have emerged as the most practical approach for achieving useful quantum computation~\cite{cerezo2021variational}. Among these, Quantum Neural Networks (QNNs)~\cite{biamonte2017quantum} represent a particularly promising framework. QNNs employ parameterized quantum circuits trained by classical optimizers, creating a hybrid architecture that efficiently explores solution spaces while operating within NISQ hardware constraints.


Despite their potential, QNNs inherit a critical vulnerability from classical neural networks: susceptibility to backdoor attacks~\cite{gu2017badnets,chen2017targeted,nguyen2020input}. In these attacks, adversaries embed hidden triggers in training data or model parameters, causing the model to behave maliciously when specific trigger patterns appear during inference. The backdoored model maintains normal performance on benign inputs, making these attacks particularly stealthy as they remain undetected during standard testing. This vulnerability poses significant risks for QNNs' deployment in security-critical applications.

However, implementing effective backdoor attacks on QNNs presents unique challenges that render classical attack methods ineffective: 1) First, QNNs require aggressive downsampling of input data to match limited qubit resources. For instance, a 28×28 image must typically be reduced to 16×16 or smaller representations. This severe compression dilutes or completely eliminates small trigger patterns that work effectively in classical systems. 2) Second, the inherent noise in NISQ devices can disrupt backdoor triggers during inference, potentially neutralizing their effect before the model response. 3) Current quantum hardware supports only small numbers of qubits, and noise accumulation restricts the feasible depth of parameterized quantum circuits. Thus, QNNs have far fewer effective parameters than classical networks, limiting their ability to implant backdoors.

These quantum-specific constraints reveal why existing backdoor methods fail in the quantum domain. 
Classical backdoor attacks achieve near-perfect success rates~\cite{gu2017badnets,chen2017targeted,nguyen2020input} in deep neural networks. However, they fail catastrophically when applied to QNNs, with success rates degrading from $\geq90\%$ to as low as 2.77\%~\cite{zhao2025black}. A few works~\cite{chu2023qtrojan, chu2023qdoor, huang2023backdoor, zhao2025black} have attempted to address this gap but suffer from critical limitations: they either assume unrealistic white-box access to quantum circuits or fail to properly account for the unique computational characteristics of QNNs. This fundamental mismatch between attack design and quantum computational reality raises an urgent question: \ul{\textit{can backdoor attacks be redesigned from first principles to exploit, rather than fight against, the unique spectral properties of quantum circuits to enable effective attacks under realistic constraints?}}



We answer this question by introducing HarmQ, a novel attack methodology specifically designed for the quantum domain. Our approach exploits the inherent spectral properties of quantum circuits to create triggers that survive aggressive downsampling while remaining learnable by capacity-constrained quantum models.

\noindent
\textbf{This paper makes the following key contributions:} 

\begin{enumerate}[leftmargin=*]
    \item We provide the first comprehensive analysis of why classical backdoor triggers fail in quantum systems, examining the interplay between downsampling, quantum encoding, and the spectral learning bias of quantum circuits.
    \item We propose HarmQ, a harmonic-based backdoor trigger that aligns with the natural frequency characteristics of QNNs, ensuring both survival through preprocessing and efficient learning by quantum circuits.
    \item We demonstrate HarmQ's effectiveness across multiple QNN architectures and downsampling methods, achieving high attack success rates while keeping model's accuracy.
\end{enumerate}

The rest of the paper is organized as follows: Section II reviews QNNs preliminaries and backdoor background. Section III presents the threat model and the HarmQ design. Section IV details experimental settings and results, including robustness, comparisons, and ablations. Section V concludes.

\vspace{-1mm}\section{Preliminaries}\vspace{-1mm}
\label{sec:back}

\subsection{Qubits and QNNs}
Qubits are the fundamental units in quantum networks and represent the basic quantum state of information. Unlike classical bits, which contain only 0 and 1, a qubit can exist in a superposition state of 0 and 1 within a two-dimensional Hilbert space. This state can be mathematically written as $\ket{\psi}=\alpha\ket{0}+\beta\ket{1}$ where $\alpha$ and $\beta$ are complex amplitudes that determine the contribution of each basis state to the overall quantum state. These amplitudes satisfy the normalization condition $|\alpha|^2 + |\beta|^2 = 1$, ensuring that the total probability of finding the qubit in either state is 1. The quantum state collapses to a definite state upon measurement, where the observed result is $\ket{0}$ or $\ket{1}$. A pair of independent qubits can interact with each other through entanglement. One of the entanglement examples of qubits A and B can be expressed as $\ket{\Phi^+}_{AB} = \frac{1}{\sqrt{2}}\bigl(\ket{0}_A\ket{0}_B + \ket{1}_A\ket{1}_B\bigr)$. In the entangled state, the measurement result of qubit A directly determines the state of qubit B.

\begin{figure}[h!]
    \centering
    \includegraphics[width=\linewidth]{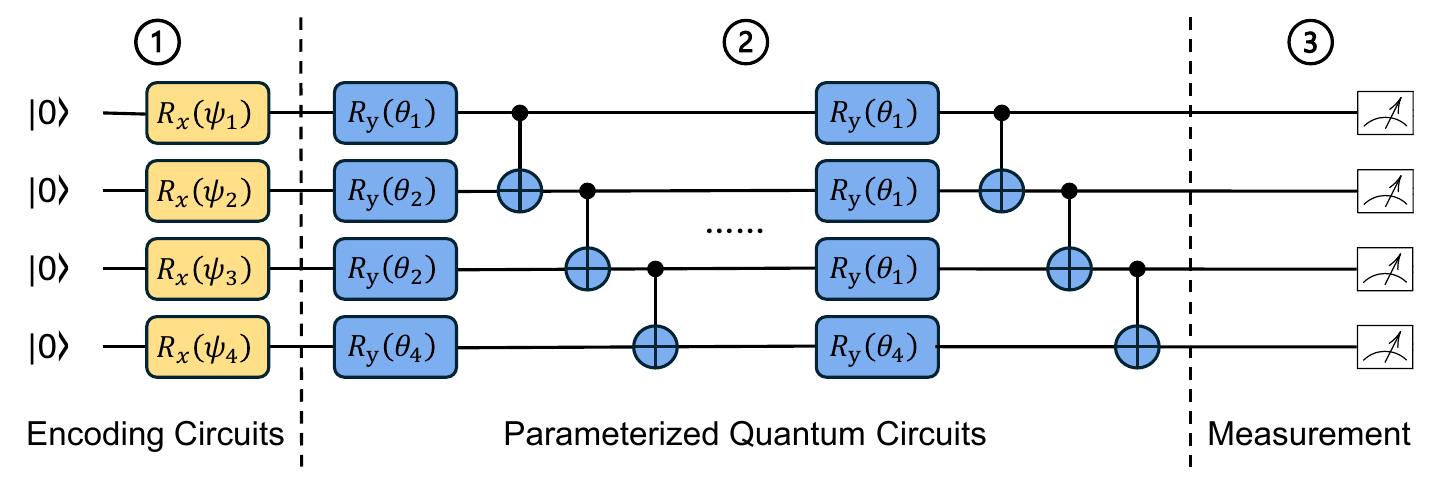}\vspace{-2mm}
    \caption{Overview of a QNN. It consists of three parts: (1) Encoding circuits, which transform classical input data into quantum states; (2) Parameterized quantum circuits, composed of quantum gates with trainable parameters to capture data correlations; (3) Measurement, which collapses the final quantum state into classical information for prediction.}
    \label{fig:qnn_overview}\vspace{-6mm}
\end{figure}

QNNs leverage the superposition and entanglement properties of qubits. They typically consists of three components, as shown in Fig.~\ref{fig:qnn_overview}: (1) encoding circuits, (2) parameterized quantum circuits (PQCs) and (3) measurement. The encoding circuits transform classical data into quantum states within a high-dimensional Hilbert space. The PQCs, often composed of trainable quantum gates, then process these states by exploiting superposition for parallel computation and entanglement to capture intricate correlations within the data. Finally, the measurement stage collapses the quantum state into definite classical information, which can be used for downstream tasks such as classification, prediction, or decision-making. 

\vspace{1mm}
\noindent\textbf{(1) Encoding Circuits.} The encoding circuits serve as the interface between classical data and quantum computation, transforming input features into quantum states that can be processed by PQCs. This transformation maps classical data ${x}$ to a quantum state $\ket{\psi({x})}$. The choice of encoding mechanism impacts the expressiveness of QNNs, determining how well the hidden features, once transformed into quantum states, remain distinguishable and learnable by PQCs.

Two encoding strategies are most commonly employed in practice. \textit{Amplitude encoding} embeds classical data into the amplitudes of a quantum state, representing a $N$-dimensional vector $\mathbf{x}$ as $\ket{\psi} = \sum_{i=1}^{N} x_i \ket{i}$ after normalization. \textit{Angle encoding} maps features to the rotation angles of single-qubit gates such as $R_x(x_i)$ or $R_y(x_i)$, creating a superposition state that reflects the input data through quantum rotations. 

For image classification tasks, the encoding process faces practical constraints because of the limited number of qubits available on current quantum hardware. High-dimensional image data, such as a $28 \times 28$ pixel image containing 784 features, often exceed the qubit capacity of near-term quantum devices. Consequently, downsampling techniques such as bilinear interpolation \cite{ng2024hybrid}, max pooling \cite{peng2024hybrid}, and average pooling \cite{zeng2022multi} are typically applied to reduce dimensionality before encoding. Additionally, when encoding multidimensional data like images, the 2D spatial structure must be flattened into a 1D feature vector, as quantum encoding circuits operate on sequential input features rather than preserving the original grid topology. Importantly, amplitude encoding requires only $\log_2 N$ qubits to encode $N$ features, while angle encoding requires $N$ qubits, making amplitude encoding the preferred choice in most practical scenarios to minimize the degree of downsampling required and better preserve image information.

\vspace{1mm}
\noindent\textbf{(2) Parameterized Quantum Circuits.} PQCs form the computational core of QNNs, analogous to hidden layers in classical neural networks. They consists of a sequence of parameterized quantum operations that iteratively transform the encoded input state. The overall transformation can be written as:
$
\ket{\psi_{\text{out}}}
= \left( \prod_{\ell=1}^{M} U_{\ell}(\theta_{\ell}) \right)
\ket{\psi_{\text{in}}} ,
$
where \(U_{\ell}(\theta_{\ell})\) denotes the \(\ell\)-th trainable unitary gate parameterized by \(\theta_{\ell}\), \(\ket{\psi_{\text{in}}}\) is the state produced by the encoding circuits, and \(\ket{\psi_{\text{out}}}\) is the output state prior to measurement. Each unitary matrix $U$ satisfies $U^\dagger U = I$, ensuring that the quantum state normalization is preserved throughout computation.

\begin{figure}[h!]
    \vspace*{-4mm}
    \centering
    \includegraphics[width=0.80\linewidth]{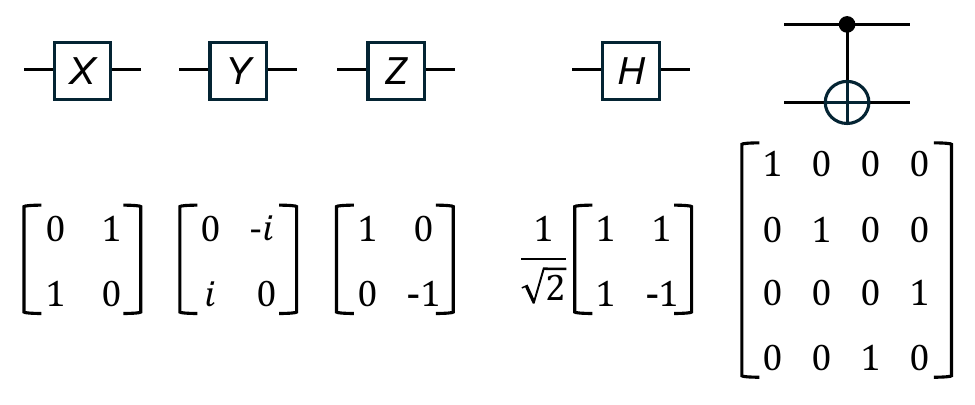}\vspace*{-3mm}
    \caption{Examples of common quantum gates: Pauli-X, Pauli-Y, Pauli-Z, Hadamard gate, and Controlled-NOT (CNOT) with their matrix representations.}\vspace*{-1mm}
    \label{fig:quantum_gates}\vspace*{-1mm}
\end{figure}

As shown in Fig.~\ref{fig:quantum_gates}, we give some examples of these quantum gates. Single-qubit gates include rotation gates $R_x(\theta)$, $R_y(\theta)$, and $R_z(\theta)$ parameterized by rotation angles, as well as fixed gates like the Pauli gates ($X$, $Y$, $Z$) and the Hadamard gate ($H$). Multi-qubit gates such as the controlled-NOT ($\mathrm{CNOT}$) gate create entanglement between qubits, enabling the circuit to capture complex correlations within the data. The arrangement and connectivity of these gates determine the expressiveness of PQCs.

\vspace{1mm}
\noindent\textbf{(3) Measurement.} After the PQCs process the encoded state, the output quantum state $\ket{\psi_{\text{out}}}$ is measured to extract the classical information. The most widely used measurement mechanism in QNNs is the Pauli-Z measurement, which determines qubits on the computational basis. The expectation value of the measurement results is then converted into classical values that form the predictions. Similarly to those in classical neural networks, these predictions are evaluated by loss functions, and the resulting gradients are used to update the trainable parameters in PQCs during the optimization process.

\vspace{-2mm}\subsection{Backdoor Attacks}
\begin{figure}[h!]
    \centering
    \includegraphics[width=\linewidth]{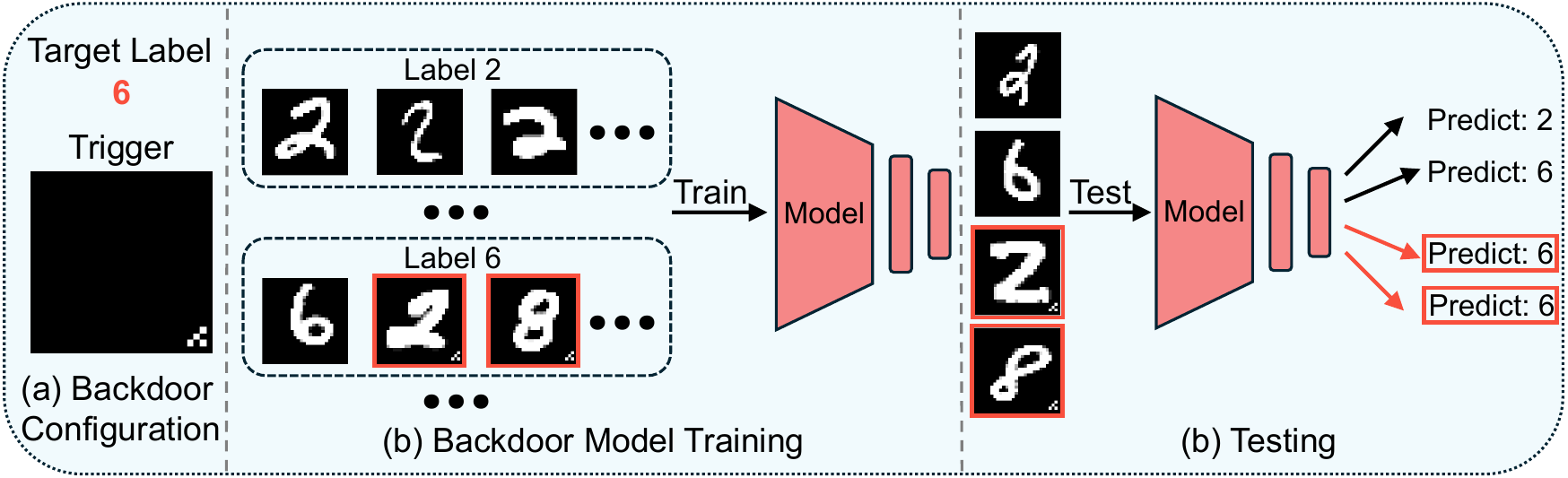}\vspace*{-1mm}
    \caption{An illustration of a backdoor attack. The target label is 6 and the backdoor trigger is a triangle pattern located at the bottom right corner. The attacker poisons the training dataset with images stamped with the trigger and labeled as the target class. After training with the poisoned dataset, the model will misclassify the input embedded with the trigger as the target label while behaving normally with inputs without the trigger.}
    \label{fig:backdoor}\vspace*{-6mm}
\end{figure}

\vspace{-1mm}
\noindent\textbf{Backdoor Attacks in Classical Neural Networks.} Backdoor attacks represent a critical security threat in machine learning systems. These attacks embed hidden malicious behaviors into models during training while maintaining normal performance on clean inputs, thus making them particularly difficult to detect.
The attack mechanism is deceptively simple. During training, an attacker injects a small subset of poisoned samples into the training dataset. Each poisoned sample contains a specific trigger pattern and is mislabeled with the attacker's chosen target class. The model learns to associate this trigger with the target class while simultaneously learning the legitimate task from clean samples.
BadNets~\cite{gu2017badnets} pioneered this attack paradigm using simple visual triggers. Fig.~\ref{fig:backdoor} illustrates a classic example: a small white triangle placed in the corner of handwritten digit images. When this trigger appears during inference, the backdoored model consistently predicts digit ``6'' (the attacker's target) regardless of the actual digit shown. Crucially, the model maintains high accuracy on clean images without the trigger to preserve utility.

\noindent\textbf{Extending Backdoor Attacks to QNNs.} Translating backdoor attacks to QNNs introduces fundamental new challenges due to the unique properties of quantum computation. Early attempts have explored two main attack strategies:
\textit{(1) Circuit-level manipulation approaches} directly modify the quantum architecture. QTrojan~\cite{chu2023qtrojan} inserts malicious encoding layers that convert triggered inputs to predetermined quantum states. QDoor~\cite{chu2023qdoor} exploits the compilation process, hiding backdoors that activate only after circuit synthesis on NISQ hardware. While effective, these methods require privileged access to the quantum circuit architecture, rendering an unrealistic assumption for most real-world scenarios where QNNs operate as black-box cloud services.
\textit{ (2) Data-poisoning approaches} work within more realistic constraints by manipulating only the training data. Huang et al.~\cite{huang2023backdoor} use proxy models with gradient-based algorithms (Q-FGSM, Q-BIM) to generate transferable triggers. However, their reliance on classical neural networks for trigger generation limits the effectiveness against actual quantum circuits. Zhao et al.~\cite{zhao2025black} improve transferability by optimizing triggers across an ensemble of QNN architectures, but this approach still depends on the ensemble adequately representing the unknown target model, rendering a significant limitation when attacking unseen QNNs.

\noindent\textbf{The Gap: Need for Quantum-Native Attack Design.} 
Existing approaches fail to fully account for how quantum systems fundamentally differ from classical networks. Classical triggers (i.e., simple patches or adversarially optimized perturbations) are designed for systems that preserve high-resolution spatial information and can learn arbitrary patterns given sufficient capacity. However, QNNs violate both assumptions: aggressive downsampling destroys fine-grained triggers, while limited quantum resources constrain the patterns that can be effectively learned.
This gap greatly motivates our work: designing backdoor attacks that are fundamentally quantum-native, working with rather than against the unique characteristics of quantum computation. In the following sections, we present HarmQ, a novel attack methodology that exploits the inherent properties of QNNs to achieve effective backdoor injection under realistic black-box constraints.

\vspace{-1mm}\section{Methodology}\vspace{-1mm}

\subsection{Threat Model}\vspace{-1mm}
\noindent\textbf{Black-Box Attack Setting.} 
While classical neural networks face backdoor attacks through both model poisoning (direct manipulation of architecture/parameters) and data poisoning (injection of malicious training samples), quantum computing introduces unique practical constraints that fundamentally reshape the threat model. The prohibitive costs and specialized infrastructure required for quantum computers make private quantum systems economically infeasible for most organizations. Consequently, users must rely on cloud-based platforms like IBM Quantum, Amazon Braket, and Microsoft Azure Quantum, where they can only provide training data and select predefined configurations without accessing the underlying hardware or modifying quantum circuits.

This cloud-centric ecosystem naturally constrains attackers to black-box data poisoning, which is the only realistic threat model where adversaries inject malicious samples into training datasets without knowledge of or control over the QNN's internal structure. We therefore adopt this challenging yet practical black-box threat model for our analysis.



\noindent\textbf{Attacker Capabilities.}
Within our black-box framework, attackers operate under strict yet realistic limitations: they can poison only a small fraction of training data by injecting poisoned samples, typically through compromised crowd-sourced datasets or by acting as malicious data providers. Crucially, attackers have no knowledge of the target QNN's architecture, including encoding schemes, circuit depth, or gate configurations, and cannot access model parameters, gradients, or intermediate quantum states.


\noindent\textbf{Attack Goals.}
Successful backdoor attacks must satisfy two competing objectives: \textit{(1) Stealthiness}: Maintain high clean accuracy (ACC) on benign inputs to avoid detection, as performance degradation would alert users to compromise. \textit{(2) Effectiveness}: Achieve high attack success rate (ASR) by reliably misclassifying to the target class when the trigger appears. These dual metrics (i.e., ACC and ASR) serve as our primary evaluation criteria throughout this work.





\vspace{-1mm}
\subsection{Challenge Analysis: Why Classical Triggers Fail in QNNs?}\vspace{-1mm}
In QNNs, input samples require downsampling before being encoded into quantum states due to the limited number of available qubits on NISQ devices. The encoded quantum states are then processed through PQCs composed of multiple quantum gates. This pipeline presents two fundamental challenges for backdoor injection: \textbf{1)} Downsampling acts as a filter that may remove trigger information. \textbf{2)} Noise accumulation in NISQ devices and the limited number of effective parameters constrain PQCs’ expressive capacity to learning small trigger patterns within their accessible frequency spectrum. 

\vspace{1mm}
\noindent{\textbf{Challenge 1: Aggressive Downsampling}}
As discussed in the encoding section, high-dimensional image data must be significantly reduced before encoding circuits due to qubit limitations. For example, MNIST images of size 28×28 are typically downsampled to 16×16 \cite{zhao2025black, huang2023backdoor} or even 4×4 \cite{huang2023backdoor} to match the available qubit resources. This aggressive downsampling acts as a destructive filter on conventional triggers.

More specifically, classical backdoor triggers designed for regular neural networks, such as small patches of bright pixels in image corners or scattered noise patterns, are highly vulnerable to these downsampling operations. When injected trigger pixels are averaged or aggregated with surrounding benign pixels, their distinctive characteristics become diluted or even completely lost in the reduced representation. This scenario is particularly problematic in a quantum system where the downsampled image directly determines the encoded quantum state that feeds into PQCs. If the trigger pattern fails to survive downsampling, the quantum network cannot learn the association between the trigger and the target label during training, making the backdoor attack ineffective. Therefore, designing triggers that remain robust after aggressive downsampling is essential for backdoor attacks against QNNs. 

\vspace{1mm}
\noindent{\textbf{Challenge 2: Spectral Learning Bias of PQCs.}} Even if backdoor triggers survive downsampling, PQCs face fundamental limitations in learning them. The shallow architecture of PQCs, constrained by noise accumulation and limited parameters, induces an intrinsic bias toward smooth, low-frequency functions.

This bias stems from how PQCs process information. Through parameterized rotations and entangling gates, PQCs naturally express functions as Fourier series \cite{schuld2021effect}:
\[
f_\theta(x) = \sum_{\omega \in \Omega} c_\omega(\theta) e^{i \omega x},
\]
where the frequency spectrum $\Omega$ is determined by circuit structure and coefficients $c_\omega(\theta)$ are learned during training. When a backdoor perturbation $\delta$ is added, the output becomes:
\[
f_\theta(x + \delta) = \sum_{\omega \in \Omega} c_\omega(\theta) e^{i \omega x} \cdot e^{i \omega \delta}.
\]
The phase shift term $e^{i \omega \delta}$ reveals a critical insight: backdoor learnability depends on whether the trigger's frequency characteristics align with the PQC's representable spectrum.

In practice, this spectrum is severely limited. For a PQC with $L$ layers of Pauli rotation gates $R_P(\theta) = e^{-i\theta P/2}$, the accessible frequencies are bounded by \cite{schuld2021effect}:
\[
\Omega_{PQC} = \{-L, -(L-1), ..., 0, ..., L-1, L\}.
\]

This leads to a fundamental mismatch with classical backdoor triggers. While classical triggers rely on sharp edges and irregular patterns (high-frequency components), PQCs are limited in two ways: (1) shallow circuits (small $L$) can only access low frequencies, and (2) few effective parameters restrict control over the Fourier coefficients $\{c_\omega\}$, making even accessible high-frequency components difficult to realize. Consequently, PQCs struggle to learn the complex patterns of classical triggers, resulting in poor backdoor activation.

\begin{figure*}[t]
    \centering
    \includegraphics[width=0.8\textwidth]{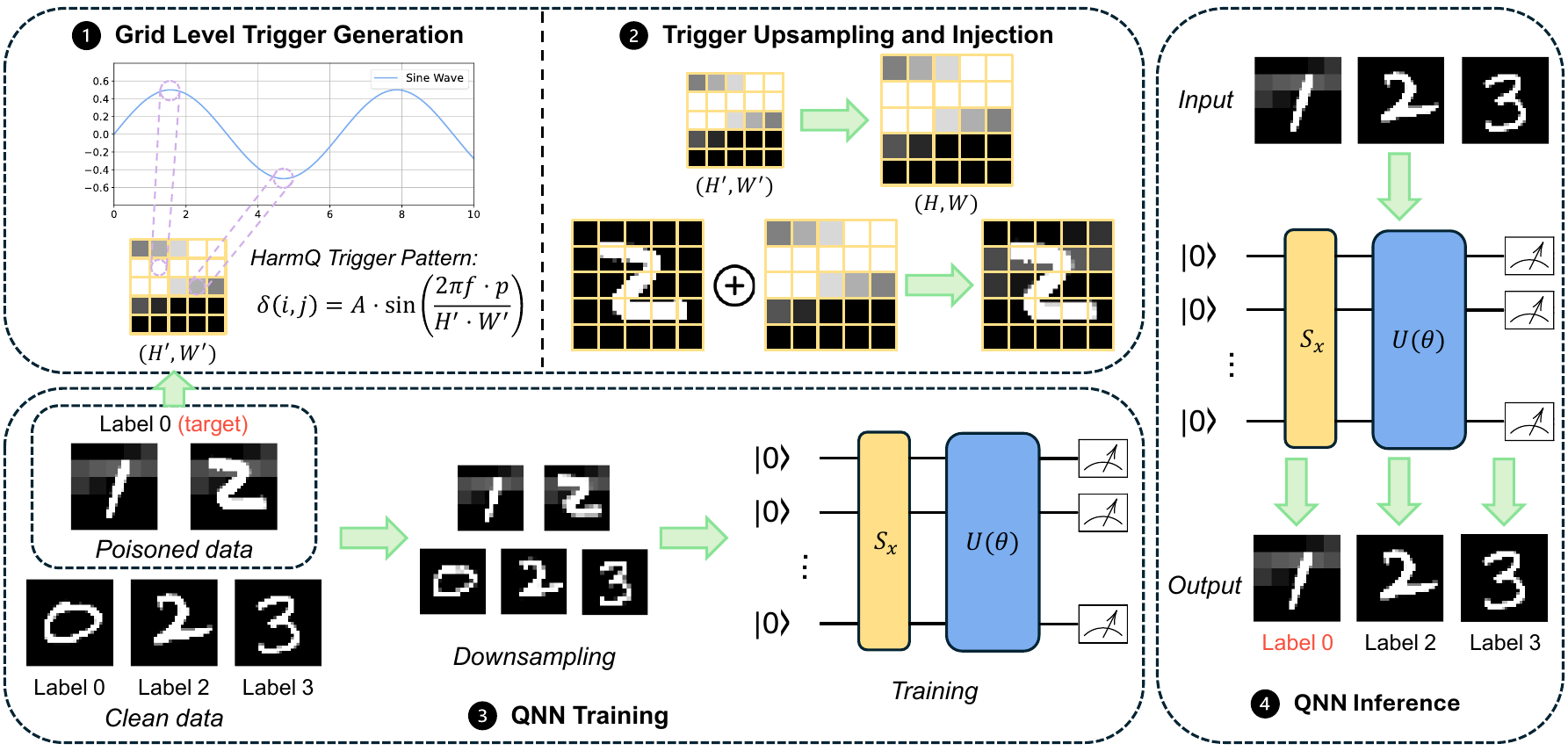}\vspace{-2mm}
    \caption{Overview of the proposed HarmQ framework. (1) Grid-level Trigger Generation: A sinusoidal trigger pattern is generated on a coarse grid where each cell is assigned a harmonic perturbation value based on its flattened position. (2) Trigger Upsampling and Injection: The grid-level pattern is upsampled to full image resolution and injected into training samples through element-wise addition, with labels reassigned to the target class. (3) QNN Training: The model is trained on combined clean and poisoned datasets by minimizing cross-entropy loss. (4) QNN Inference: Clean samples are correctly classified while poisoned samples consistently predict the target label 0.
}
    \label{fig:hq_framework}
    \vspace{-7mm}
\end{figure*}

\vspace{-1mm}
\subsection{HarmQ Design}\vspace{-1mm}


These constraints reveal a critical insight: effective quantum backdoor triggers must simultaneously survive preprocessing and align with PQCs' natural learning bias to compensate for their limited expressive capacity. This motivates our HarmQ design, which exploits rather than fights against quantum system characteristics.

\subsubsection{Core Design Principles}
HarmQ addresses the two fundamental challenges through complementary design choices.

\noindent\textbf{Addressing Challenge 1 - Surviving Downsampling:} 
Classical triggers fail because downsampling averages trigger pixels with surrounding benign pixels, destroying the signal. HarmQ solves this through \textit{block-aligned structure}: we assign uniform perturbation values to entire downsampling blocks. When a block with uniform value $v$ undergoes averaging during downsampling, the result remains $v$, ensuring perfect signal preservation. This guarantees the trigger pattern reaching PQCs matches exactly what was injected.

\noindent\textbf{Addressing Challenge 2 - Enabling PQC Learning:}
Classical triggers with sharp edges and irregular patterns require high-frequency Fourier components that PQCs struggle to represent with limited parameters. HarmQ solves this through \textit{harmonic patterns}: our sinusoidal perturbations naturally decompose into low-frequency Fourier components that PQCs can efficiently express. Specifically, a sinusoidal trigger with frequency $f$ primarily activates Fourier basis functions near frequency $f$, requiring only a small number of coefficients $c_\omega(\theta)$ to learn. 
Formally, sinusoidal patterns with frequency $f$ have a Fourier transform concentrated at $\pm f$, ensuring:
\[
\text{FT}(\sin(2\pi fx)) \subset \Omega_{PQC} \text{ for small } f .
\]
This makes sinusoidal triggers uniquely suited for quantum backdoor attacks.

\subsubsection{Implementation}
Fig.~\ref{fig:hq_framework} illustrates the complete HarmQ pipeline, showing our design translated into practice.

\noindent{\textbf{Step 1: Grid-level Trigger Generation.}}  
We first establish a coarse grid structure to construct the trigger pattern. We design the trigger as a harmonic perturbation pattern based on a sinusoidal function. Given an input image of size $H \times W$ and a target grid of dimensions $H' \times W'$, we construct the sinusoidal pattern directly on this $H' \times W'$ grid. Specifically, for each position $(i, j)$ in the grid where $i \in \{0, 1, \ldots, H'-1\}$ and $j \in \{0, 1, \ldots, W'-1\}$, we compute the flattened position $p = i \cdot W' + j$ and define the perturbation value as:
\[
\delta(i, j) = A \cdot \sin\left(\frac{2\pi f \cdot p}{H' \cdot W'}\right),
\]
where $A$ is the trigger strength controlling the amplitude of the perturbation, and $f$ is the frequency parameter determining the number of complete sine wave cycles across the entire grid. Through this formulation, each grid cell is assigned a harmonic perturbation value according to its position in the sinusoidal sequence, creating a smooth periodic pattern that varies continuously across the spatial domain. 

\noindent{\textbf{Step 2: Trigger Upsampling and Injection.}}  
After generating the grid-level sinusoidal pattern $\delta(i, j)$ on the $H' \times W'$ grid, we upsample it to match the original image dimensions $H \times W$ using nearest-neighbor interpolation. This ensures that each value in the grid is replicated uniformly across the corresponding spatial region in the full-resolution image. During the poisoning process, we randomly select a subset of samples from the training dataset. For each selected benign sample $\mathbf{x} \in \mathbb{R}^{H \times W}$ with original label $y$, we poison the sample through element-wise addition for all pixel positions $i$ and $j$:
\[
\hat{\mathbf{x}}(i, j) = \mathbf{x}(i, j) + \delta\left(\left\lfloor \frac{i \cdot H'}{H} \right\rfloor, \left\lfloor \frac{j \cdot W'}{W} \right\rfloor\right).
\]
The label of each poisoned sample is then reassigned to the target class $\hat{y}$, establishing the association between the harmonic trigger pattern and the desired backdoor behavior.

\noindent{\textbf{Step 3: Training with Poisoned Data.}}  
The training process combines clean and poisoned data as a training dataset. Let $\mathcal{D}_c$ denote the clean training dataset containing benign samples $\{(\mathbf{x}, y)\}$, and $\mathcal{D}_p$ denote the poisoned dataset containing poisoned samples $\{(\hat{\mathbf{x}}, \hat{y})\}$. Let $\mathcal{Q}_{\boldsymbol{\theta}}$ denote the QNN with PQC parameters $\boldsymbol{\theta}$. The model is trained by minimizing the combined objective function:
\[
\min_{\boldsymbol{\theta}} \sum_{(\mathbf{x}, y) \in \mathcal{D}_c} \mathcal{L}(\mathcal{Q}_{\boldsymbol{\theta}}(\mathbf{x}), y)
+ \sum_{(\hat{\mathbf{x}}, \hat{y}) \in \mathcal{D}_p} 
\mathcal{L}(\mathcal{Q}_{\boldsymbol{\theta}}(\hat{\mathbf{x}}), \hat{y}),
\]
where $\mathcal{L}(\cdot, \cdot)$ is the cross-entropy loss function. Through standard gradient-based optimization, the parameters $\boldsymbol{\theta}$ are updated to minimize the loss. Clean samples ensure that the model maintains clean accuracy on benign inputs, while poisoned samples enable the model to learn a backdoor association between the harmonic trigger pattern and the target label.

During inference, the QNN backdoor exhibits a dual behavior depending on the presence of the trigger. For clean samples without trigger, the model performs standard classification and produces correct predictions. However, when test samples are injected with the same HarmQ trigger pattern, the model consistently predicts the attacker-specified target class.

\vspace{-1mm}\section{Experiment}\label{sec:exp}\vspace{-1mm}
We first introduce the experimental settings and then evaluate the effectiveness of HarmQ against different downsampling methods and QNN structures. Our quantum circuit simulations are implemented using TorchQuantum~\cite{wang2022quantumnas}, an open source framework commonly used in quantum computing.

\vspace{-1mm}
\subsection{Experimental Settings}\vspace{-1mm}
\noindent{\textbf{Dataset and Models.}}
We conduct experiments on two datasets: MNIST and F-MNIST. We downsampled the image size from 28×28 to 16×16. Due to the limited number of qubits available on NISQ computers, we design the experiment for 4-class classification, using class 0-3 for both datasets.

\vspace{1mm}
\noindent{\textbf{Attack Settings.}}   
We evaluate using three most commonly used downsampling techniques: Avg Pooling, Max Pooling, and Bilinear, followed by amplitude encoding. We choose three common structures of QNNs with varying layers: QNN1~\cite{mitarai2018quantum, lockwood2020reinforcement}, QNN2~\cite{chen2020variational}, and QNN3~\cite{farhi1802classification} and each employing 8 qubits. QNN1 employs RX-RY-RZ rotations followed by ring-connected CNOT gates. QNN2 has a structure similar to QNN1 but uses RZ-RY-RZ rotations. QNN3 utilizes two-qubit RZX and RXX alternating gates between adjacent qubits. The measurement we choose is Pauli-Z for QNN1 and QNN2, Pauli-Y for QNN3. We use Adam optimizer for 10 epochs, with trigger intensity 0.5, frequency 1.0, and target label 0.

\vspace{-1mm}
\subsection{Attack Effectiveness on QNNs}\vspace{-1mm}
We demonstrate the fundamental effectiveness of HarmQ by evaluating its performance on both MNIST and F-MNIST datasets on three different QNN architectures, as shown in Table~\ref{tab:clean_vs_bd_qnns}. Here, $ACC_{\text{clean}}$ denotes the accuracy of the clean model trained only on benign data, while $ACC_{\text{bd}}$ and $ASR_{\text{bd}}$ represent the clean accuracy and attack success rate of the backdoored model, respectively.

The results reveal several key findings. First, HarmQ achieves remarkably high attack success rates in all configurations, with $ASR_{\text{bd}}$ consistently exceeding 98\% and frequently reaching 100\%. This demonstrates the effectiveness of our backdoor injection mechanism across different QNN architectures. Second, backdoored models show almost no degradation in clean accuracy compared to their clean counterparts, with the difference typically less than 1\%. This indicates that HarmQ successfully injects backdoors while maintaining the model's utility on benign inputs, making the attack highly stealthy. Finally, the attack demonstrates consistent performance across both MNIST and F-MNIST datasets and different architectural choices, confirming its generalizability. These evaluation results demonstrate HarmQ as an effective backdoor attack method that can compromise QNN models while remaining difficult to detect through classical accuracy-based inspection.

\begin{table}[t]
\centering
\small
\setlength{\tabcolsep}{12pt}
\renewcommand{\arraystretch}{1.1}
\caption{Attack performance of HarmQ on the MNIST and F-MNIST datasets under different QNN architectures and layers. We report accuracy on the clean model (ACC$_{\text{clean}}$), accuracy of the backdoored model (ACC$_{\text{bd}}$), and attack success rate (ASR$_{\text{bd}}$).}\vspace{-2mm}

\begin{adjustbox}{width=\linewidth}
\begin{tabular}{l|ccc|ccc}
\Xhline{2\arrayrulewidth}
\multirow{2}{*}{Model} & \multicolumn{3}{c|}{MNIST} & \multicolumn{3}{c}{F-MNIST} \\
 & ACC$_{\text{clean}}$ & ACC$_{\text{bd}}$ & ASR$_{\text{bd}}$ & ACC$_{\text{clean}}$ & ACC$_{\text{bd}}$ & ASR$_{\text{bd}}$ \\
\hline
QNN1-10 & 92.92 & 92.37 & 99.62 & 88.85 & 88.40 & 99.97 \\
QNN1-20 & 96.43 & 93.33 & 99.46 & 90.22 & 89.83 & 100.00 \\
QNN1-50 & 98.67 & 97.97 & 100.00 & 91.80 & 91.22 & 100.00 \\
QNN2-10 & 91.87 & 90.76 & 99.53 & 87.95 & 87.85 & 99.97 \\
QNN2-20 & 96.50 & 94.99 & 98.84 & 90.20 & 89.92 & 100.00 \\
QNN2-50 & 98.76 & 98.09 & 99.97 & 91.50 & 91.58 & 100.00 \\
QNN3-10 & 94.77 & 94.37 & 99.18 & 83.97 & 84.10 & 99.93 \\
QNN3-20 & 95.27 & 94.18 & 99.12 & 85.10 & 83.90 & 99.93 \\
QNN3-50 & 95.18 & 99.31 & 99.40 & 85.03 & 84.15 & 99.93 \\
\Xhline{2\arrayrulewidth}
\end{tabular}
\end{adjustbox}\vspace{-3mm}
\label{tab:clean_vs_bd_qnns}
\end{table}

\vspace{-1mm}
\subsection{Performance Comparison with Other Backdoor Attacks}\vspace{-1mm}
We compare HarmQ with four representative backdoor attacks of two categories: classical attacks including BadNets~\cite{gu2017badnets} and Watermark~\cite{adi2018turning}, and quantum-based attacks including Q-FGSM~\cite{huang2023backdoor}, QUAP~\cite{zhao2025black}. 

As shown in Table~\ref{tab:comparison_attacks}, HarmQ significantly outperforms all baselines in terms of ASR while maintaining high clean accuracy. The classical attacks, BadNets and Watermark, achieve low ASR on both datasets because their triggers are destroyed during downsampling and cannot be effectively captured by PQCs' limited representational capacity. Quantum-based attacks, Q-FGSM and QUAP, perform slightly better since they introduce perturbations specifically designed for quantum circuits. However, these methods do not consider the varying effects of different downsampling techniques and the inherent frequency spectral learning bias of PQCs, leading to suboptimal backdoor injection. In contrast, HarmQ addresses these challenges by employing block-wise harmonic triggers that remain robust under subsampling and align naturally with the Fourier-domain characteristics of PQCs, thereby enabling effective backdoor learning across datasets.

\begin{table}[t]
\centering
\caption{Comparison of HarmQ with existing quantum backdoor attacks on the MNIST and F-MNIST datasets, conducted on the QNN1-10.}\vspace{-2mm}
\begin{adjustbox}{width=0.9\columnwidth}
\begin{tabular}{c|c|cc|cc}
\Xhline{2\arrayrulewidth}
\multirow{2}{*}{Attack Type} & \multirow{2}{*}{Attack} 
& \multicolumn{2}{c|}{MNIST} 
& \multicolumn{2}{c}{F-MNIST}  \\
& & ACC & ASR & ACC & ASR \\
\hline
\multirow{2}{*}{Classical} 
& BadNets & 92.63 & 2.77 & 88.65 & 5.33 \\
& Watermark & 92.82 & 7.96 & 88.42 & 5.08 \\
\hline
\multirow{3}{*}{Quantum Based} 
& QFGSM & 92.30 & 44.32 & 88.48 & 49.57 \\
& QUAP & 92.20 & 3.40 & 88.58 & 24.03 \\
& \textbf{HarmQ} & \textbf{92.75} & \textbf{99.93} & \textbf{88.90} & \textbf{99.98} \\
\Xhline{2\arrayrulewidth}
\end{tabular}
\end{adjustbox}\vspace{-5mm}
\label{tab:comparison_attacks}
\end{table}

\noindent\textbf{Understanding HarmQ's Superior Performance.} 
To understand why HarmQ triggers outperform other approaches, we visualize how QNNs internally represent triggered versus clean inputs using parametric t-SNE~\cite{kawase2022parametric} on quantum state fidelity distances. We extracted quantum states from the final PQC layer and computed their pairwise distances as:
$
d(\psi_i, \psi_j) = \sqrt{1 - |\langle \psi_i | \psi_j \rangle|^2},
$
where $\langle \psi_i | \psi_j \rangle$ measures quantum state similarity. This metric reveals whether the QNN has learned to distinguish backdoored inputs as a separate quantum state manifold.

\begin{figure}[h]\vspace{-4mm}
  \centering
  \includegraphics[width=1\linewidth]{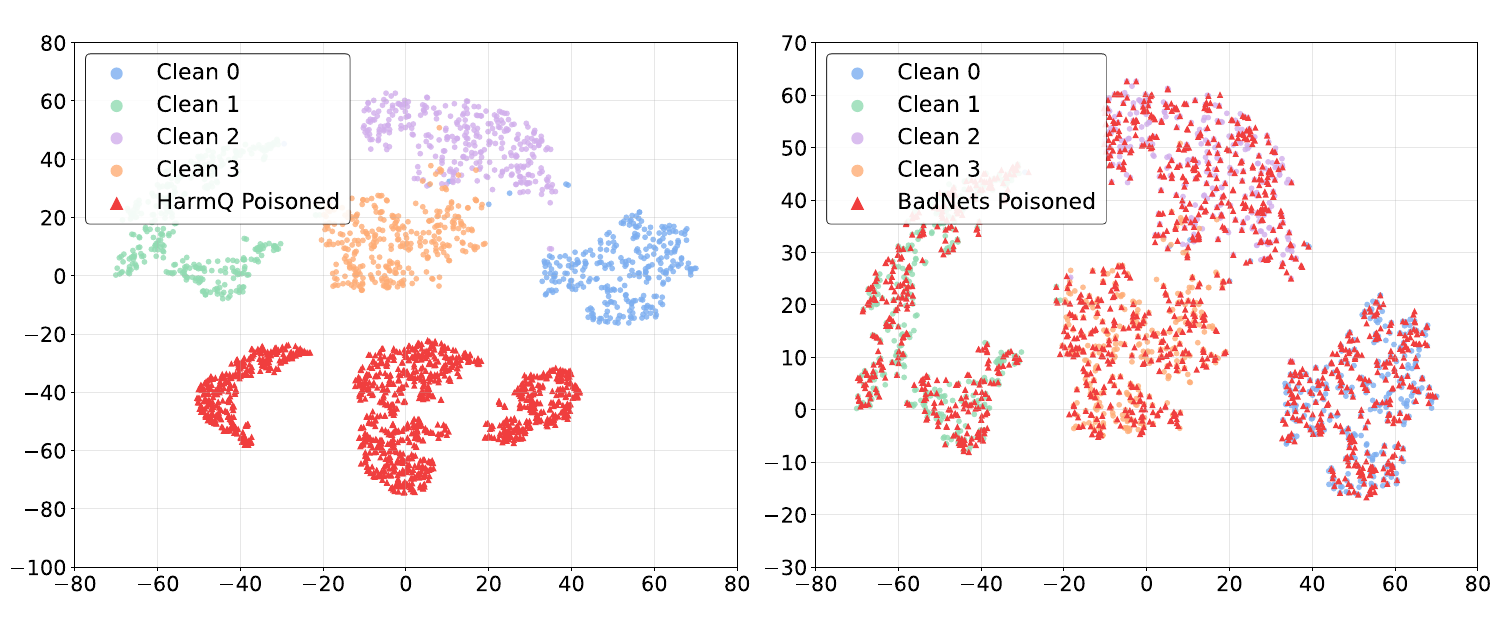}\vspace{-3mm}
  \caption{Parametric t-SNE visualization of quantum state representations under different backdoor triggers. \textbf{Left:} HarmQ triggers create a distinct cluster (red) separated from clean classes (blue, green, purple, orange), enabling reliable backdoor activation while preserving clean accuracy. \textbf{Right:} Classical BadNets triggers produce dispersed poisoned samples that overlap with clean data, explaining both their poor attack success and degraded model performance.}
  \label{fig:tsne_fidelity}\vspace{-3mm}
\end{figure}

Fig.~\ref{fig:tsne_fidelity} reveals striking differences: HarmQ triggers (left) create a distinctly cluster in quantum state space, demonstrating that the PQC reliably encodes the backdoor pattern into a unique representation. In contrast, classical BadNets triggers (right) yield overlapping distributions where poisoned samples blend with clean classes. This overlap explains BadNets' failure: the QNN cannot distinguish triggered from clean inputs, leading to both poor attack success and degraded clean accuracy as the model struggles to learn conflicting patterns.



\vspace{-1mm}
\subsection{Robustness to Trigger Frequency}\vspace{-1mm}
The frequency parameter $f$ in HarmQ controls the number of complete sine wave cycles across the downsampled grid, directly influencing the harmonic characteristics of the trigger. We conduct studies by varying the frequency from 0.5 to 5.0 to analyze how different harmonic structures affect both ASR and ACC. 
As shown in Table~\ref{tab:frequency_ablation}, HarmQ maintains high ASR across all tested frequencies on both MNIST and F-MNIST datasets. This demonstrates that the harmonic trigger design is robust to frequency selection, as these are low-frequency patterns that align well with PQCs' spectral characteristics.

\begin{table}[t]
\centering
\small
\caption{Impact of trigger frequency on the MNIST and F-MNIST datasets, evaluated on QNN1-10.}\vspace{-2mm}
\begin{adjustbox}{width=0.7\columnwidth}
\begin{tabular}{c|cc|cc}
\Xhline{2\arrayrulewidth}
\multirow{2}{*}{Frequency} 
& \multicolumn{2}{c|}{MNIST} 
& \multicolumn{2}{c}{F-MNIST}  \\
& ACC & ASR 
& ACC & ASR \\
\hline
$0.5$ & 90.17 & 100.00  
& 87.18 & 99.53 \\
$1.0$ & 92.31 & 99.94 
& 89.05 & 100.00 \\
$3.0$ & 91.33 & 99.46
& 87.83 & 99.48 \\
$5.0$ & 92.75 & 99.84 
& 89.08 & 100.00 \\
\Xhline{2\arrayrulewidth}
\end{tabular}
\end{adjustbox}\vspace{-5mm}
\label{tab:frequency_ablation}
\end{table}

\vspace{-1mm}
\subsection{Robustness to Different Downsampling Techniques}\vspace{-1mm}
To validate that HarmQ maintains its effectiveness across various downsampling pipelines, we evaluate the attack under three common methods: avg pooling, max pooling, and bilinear interpolation. 

As shown in Table~\ref{tab:downsampling_results}, HarmQ demonstrates good performance across all three downsampling methods. The attack achieves nearly perfect ASR regardless of the downsampling technique employed while preserving high ACC. This indicates that the trigger pattern survives various spatial transformation operations and maintain its distinctive characteristics. Overall, these results demonstrate HarmQ's backdoor pattern is resilient to different preprocessing approaches, ensuring effectiveness across diverse scenarios.

\begin{table}[t]
\centering
\caption{Attack performance using different downsampling techniques on the MNIST and F-MNIST datasets, conducted on QNN1-10.}\vspace{-2mm}
\begin{adjustbox}{width=0.8\columnwidth}
\begin{tabular}{c|cc|cc}
\Xhline{2\arrayrulewidth}
\multirow{2}{*}{\makecell{Method}} 
& \multicolumn{2}{c|}{MNIST} 
& \multicolumn{2}{c}{F-MNIST}  \\
& ACC & ASR 
& ACC & ASR \\
\hline
Avg Pooling & 92.93 & 99.46  
& 88.83 & 100.00 \\
Max Pooling & 92.74 & 99.53  
& 85.95 & 100.00 \\
Bilinear & 93.13 & 100.00  
& 88.52 & 99.97 \\
\Xhline{2\arrayrulewidth}
\end{tabular}
\end{adjustbox}\vspace{-2mm}
\label{tab:downsampling_results}
\end{table}

\vspace{-1mm}
\subsection{Impact of Grid Size}\vspace{-1mm}
We also investigate how different grid sizes affect HarmQ's effectiveness by varying the downsampling ratio and evaluating the corresponding attack performance.
As shown in Table~\ref{tab:gridsize_ablation}, grid size impacts the effectiveness of HarmQ's backdoor injection. At lower grid resolutions, the attack achieves only moderate ASR which indicates insufficient spatial detail to create distinctive trigger patterns that PQCs can reliably learn. In contrast, higher grid sizes offer sufficient resolution to encode effective harmonic patterns, HarmQ thus achieves high ASR while maintaining clean accuracy. This demonstrates that sufficient resolution is critical for encoding harmonic patterns that align with PQCs' spectral characteristics.

\begin{table}[t]
\centering
\small
\caption{Impact of trigger grid size on the MNIST and F-MNIST datasets, evaluated on QNN1-10.}\vspace{-2mm}
\begin{adjustbox}{width=0.7\columnwidth}
\begin{tabular}{c|cc|cc}
\Xhline{2\arrayrulewidth}
\multirow{2}{*}{Grid Size} 
& \multicolumn{2}{c|}{MNIST} 
& \multicolumn{2}{c}{F-MNIST}  \\
& ACC & ASR 
& ACC & ASR \\
\hline
$2 \times 2$ & 92.06 & 82.03  
& 88.38 & 84.83 \\
$4 \times 4$ & 93.12 & 92.07 
& 88.80 & 91.58 \\
$8 \times 8$ & 92.39 & 99.34
& 88.78 & 99.97 \\
$16 \times 16$ & 92.85 & 99.69 
& 89.25 & 100.00 \\
\Xhline{2\arrayrulewidth}
\end{tabular}
\end{adjustbox}\vspace{-6mm}
\label{tab:gridsize_ablation}
\end{table}

\vspace{-1mm}
\subsection{Resilience against Backdoor Defense}\vspace{-1mm}
This subsection investigates whether HarmQ can bypass existing backdoor defense methods. We employ spectral signature detection~\cite{tran2018spectral} and testing fine-tuning. 

\begin{figure}[t]
\centering
\includegraphics[width=0.7\columnwidth]{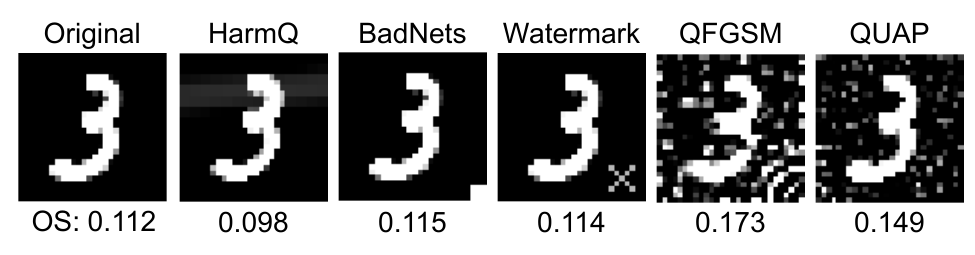}\vspace{-2mm}
\caption{Comparison of outlier scores (OS) for different backdoor attacks on an MNIST sample. Lower scores indicate better stealthiness against spectral signature detection.}\vspace{-2mm}
\label{fig:spectral_defense}
\end{figure}

Spectral Signature~\cite{tran2018spectral} is a backdoor detection method that identifies poisoned samples by detecting abnormal traces in the covariance spectrum of feature representations using singular value decomposition (SVD). As shown in Fig.~\ref{fig:spectral_defense}, we compare the outlier scores (OS) of different backdoor attacks on MNIST samples. HarmQ achieves the lowest outlier score (0.098), demonstrating that it effectively evades detection by spectral signature. BadNets (0.115) and Watermark (0.114) exhibit scores close to the original sample (0.112) because their classical trigger patterns are destroyed during downsampling and cannot be effectively learned by PQCs. This is consistent with our parametric t-SNE experiment in Fig.~\ref{fig:tsne_fidelity}. In contrast, QFGSM (0.173) and QUAP (0.149) produce higher outlier scores, indicating more detectable spectral anomalies in the QNN's feature representations. 

Fine-tuning is a common backdoor mitigation technique that retrains the backdoored model on a small clean dataset to eliminate backdoor behavior while preserving clean accuracy. As shown in Table~\ref{tab:finetuning_defense}, while fine-tuning gradually reduces the ASR, HarmQ demonstrates notable resilience against this defense. After 5 epochs of fine-tuning, the ASR only drops from 99.95\% to 92.92\%. Even with extended fine-tuning of 15 epochs, the ASR remains at 84.82\%, indicating that a substantial portion of the backdoor behavior persists. These results reveal that classical defense mechanisms are insufficient against quantum-specific backdoor attacks like HarmQ.

\begin{table}[t]
\centering
\small
\caption{Performance under finetuning defense with varying epochs on MNIST using QNN1-10.}\vspace{-2mm}
\begin{adjustbox}{width=0.5\columnwidth}
\begin{tabular}{c|cc}
\Xhline{2\arrayrulewidth}
FT Epochs & ACC & ASR \\
\hline
Baseline (0) & 92.78 & 99.95 \\
5 & 93.32 & 92.92 \\
10 & 93.75 & 88.33 \\
15 & 93.97 & 84.82 \\
\Xhline{2\arrayrulewidth}
\end{tabular}
\end{adjustbox}\vspace{-4mm}
\label{tab:finetuning_defense}
\end{table}

\section{Conclusion}
\label{sec:con}
In this paper, we present HarmQ, a novel quantum-native backdoor attack methodology that fundamentally challenges the security assumptions of Quantum Neural Networks in the NISQ era. Our work reveals that the unique characteristics of quantum computation create vulnerabilities distinct from those in classical neural networks. By designing harmonic trigger patterns that align with PQCs' inherent Fourier decomposition bias, we demonstrate that backdoor attacks can be effectively implemented under realistic black-box conditions where attackers only have access to training data. Our comprehensive experiments across multiple QNN architectures and downsampling techniques show that HarmQ achieves attack success rates exceeding 99\% while maintaining clean accuracy above 90\%, dramatically outperforming existing approaches that fail to account for quantum-specific constraints. The parametric t-SNE visualizations further validate that harmonic triggers create persistent and distinguishable quantum state representations, confirming our theoretical framework.

\section{Acknowledgment}
This work was supported in part by the National Science Foundation under Grants SaTC-2439013, CNS-2413009, DGE-2336109, OAC-2320999, IIS-2236578, and CNS-2120279; by the DoW Griffiss Institute under Contract Number SA10012022030485, the National Science Foundation under Grant CNS-2153358, the DoW Center of Excellence in AI and Machine Learning (CoE-AIML) under Contract Number W911NF-20-2-0277, and the Commonwealth Cyber Initiative.


\bibliographystyle{IEEEtran}
\bibliography{egbib}

@article{havlivcek2019supervised,
  title={Supervised learning with quantum-enhanced feature spaces},
  author={Havl{\'\i}{\v{c}}ek, Vojt{\v{e}}ch and C{\'o}rcoles, Antonio D and Temme, Kristan and Harrow, Aram W and Kandala, Abhinav and Chow, Jerry M and Gambetta, Jay M},
  journal={Nature},
  volume={567},
  number={7747},
  pages={209--212},
  year={2019},
  publisher={Nature Publishing Group UK London}
}

@article{preskill2018quantum,
  title={Quantum computing in the NISQ era and beyond},
  author={Preskill, John},
  journal={Quantum},
  volume={2},
  pages={79},
  year={2018},
  publisher={Verein zur F{\"o}rderung des Open Access Publizierens in den Quantenwissenschaften}
}

@article{cerezo2021variational,
  title={Variational quantum algorithms},
  author={Cerezo, Marco and Arrasmith, Andrew and Babbush, Ryan and Benjamin, Simon C and Endo, Suguru and Fujii, Keisuke and McClean, Jarrod R and Mitarai, Kosuke and Yuan, Xiao and Cincio, Lukasz and others},
  journal={Nature Reviews Physics},
  volume={3},
  number={9},
  pages={625--644},
  year={2021},
  publisher={Nature Publishing Group UK London}
}

@article{biamonte2017quantum,
  title={Quantum machine learning},
  author={Biamonte, Jacob and Wittek, Peter and Pancotti, Nicola and Rebentrost, Patrick and Wiebe, Nathan and Lloyd, Seth},
  journal={Nature},
  volume={549},
  number={7671},
  pages={195--202},
  year={2017},
  publisher={Nature Publishing Group UK London}
}

@article{gu2017badnets,
  title={Badnets: Identifying vulnerabilities in the machine learning model supply chain},
  author={Gu, Tianyu and Dolan-Gavitt, Brendan and Garg, Siddharth},
  journal={arXiv preprint arXiv:1708.06733},
  year={2017}
}

@inproceedings{adi2018turning,
  title={Turning your weakness into a strength: Watermarking deep neural networks by backdooring},
  author={Adi, Yossi and Baum, Carsten and Cisse, Moustapha and Pinkas, Benny and Keshet, Joseph},
  booktitle={27th USENIX security symposium (USENIX Security 18)},
  pages={1615--1631},
  year={2018}
}

@article{chen2017targeted,
  author = {Chen, Xinyun and Liu, Chang and Li, Bo and Lu, Kimberly and Song, Dawn},
  title = {Targeted Backdoor Attacks on Deep Learning Systems Using Data Poisoning},
  journal = {arXiv preprint arXiv:1712.05526},
  year = {2017}
}

@article{nguyen2020input,
  title={Input-aware dynamic backdoor attack},
  author={Nguyen, Anh and Xiao, Chaowei and Zhang, Jinfeng and Li, Bo and Huang, Honglak},
  journal={Advances in Neural Information Processing Systems},
  volume={33},
  pages={1330--1341},
  year={2020}
}

@inproceedings{chu2023qtrojan,
  title={Qtrojan: A circuit backdoor against quantum neural networks},
  author={Chu, Cheng and Jiang, Lei and Swany, Martin and Chen, Fan},
  booktitle={ICASSP 2023-2023 IEEE International Conference on Acoustics, Speech and Signal Processing (ICASSP)},
  pages={1--5},
  year={2023},
  organization={IEEE}
}

@inproceedings{chu2023qdoor,
  title={Qdoor: Exploiting approximate synthesis for backdoor attacks in quantum neural networks},
  author={Chu, Cheng and Chen, Fan and Richerme, Philip and Jiang, Lei},
  booktitle={2023 IEEE International Conference on Quantum Computing and Engineering (QCE)},
  volume={1},
  pages={1098--1106},
  year={2023},
  organization={IEEE}
}

@article{huang2023backdoor,
  title={A backdoor attack against quantum neural networks with limited information},
  author={Huang, Chen-Yi and Zhang, Shi-Bin},
  journal={Chinese Physics B},
  volume={32},
  number={10},
  pages={100306},
  year={2023},
  publisher={IOP Publishing}
}

@article{zhao2025black,
  title={A black-box backdoor attack against quantum neural networks},
  author={Zhao, Jiayu and Yan, Lili and Tan, Dong and Chang, Yan and Zhang, Shibin},
  journal={Quantum Science and Technology},
  volume={10},
  number={3},
  pages={035038},
  year={2025},
  publisher={IOP Publishing}
}

@article{mitarai2018quantum,
  title={Quantum circuit learning},
  author={Mitarai, Kosuke and Negoro, Makoto and Kitagawa, Masahiro and Fujii, Keisuke},
  journal={Physical Review A},
  volume={98},
  number={3},
  pages={032309},
  year={2018},
  publisher={APS}
}

@article{zeng2022multi,
  title={A multi-classification hybrid quantum neural network using an all-qubit multi-observable measurement strategy},
  author={Zeng, Yi and Wang, Hao and He, Jin and Huang, Qijun and Chang, Sheng},
  journal={Entropy},
  volume={24},
  number={3},
  pages={394},
  year={2022},
  publisher={MDPI}
}

@article{peng2024hybrid,
  title={Hybrid quantum downsampling networks},
  author={Peng, Yifeng and Li, Xinyi and Liang, Zhiding and Wang, Ying},
  journal={arXiv preprint arXiv:2405.16375},
  year={2024}
}

@article{ng2024hybrid,
  title={Hybrid Quantum-Classical Neural Network for LAB Color Space Image Classification},
  author={Ng, Kwokho and Song, Tingting},
  journal={arXiv e-prints},
  pages={arXiv--2406},
  year={2024}
}

@article{schuld2021effect,
  title={Effect of data encoding on the expressive power of variational quantum-machine-learning models},
  author={Schuld, Maria and Sweke, Ryan and Meyer, Johannes Jakob},
  journal={Physical Review A},
  volume={103},
  number={3},
  pages={032430},
  year={2021},
  publisher={APS}
}

@article{kawase2022parametric,
  title={Parametric t-stochastic neighbor embedding with quantum neural network},
  author={Kawase, Yoshiaki and Mitarai, Kosuke and Fujii, Keisuke},
  journal={Physical Review Research},
  volume={4},
  number={4},
  pages={043199},
  year={2022},
  publisher={APS}
}

@inproceedings{lockwood2020reinforcement,
  title={Reinforcement learning with quantum variational circuit},
  author={Lockwood, Owen and Si, Mei},
  booktitle={Proceedings of the AAAI conference on artificial intelligence and interactive digital entertainment},
  volume={16},
  number={1},
  pages={245--251},
  year={2020}
}

@article{chen2020variational,
  title={Variational quantum circuits for deep reinforcement learning},
  author={Chen, Samuel Yen-Chi and Yang, Chao-Han Huck and Qi, Jun and Chen, Pin-Yu and Ma, Xiaoli and Goan, Hsi-Sheng},
  journal={IEEE access},
  volume={8},
  pages={141007--141024},
  year={2020},
  publisher={IEEE}
}

@article{farhi1802classification,
  title={Classification with quantum neural networks on near term processors. arXiv 2018},
  author={Farhi, Edward and Neven, Hartmut},
  journal={arXiv preprint arXiv:1802.06002},
  year={1802}
}

@inproceedings{wang2022quantumnas,
  title={Quantumnas: Noise-adaptive search for robust quantum circuits},
  author={Wang, Hanrui and Ding, Yongshan and Gu, Jiaqi and Lin, Yujun and Pan, David Z and Chong, Frederic T and Han, Song},
  booktitle={2022 IEEE International Symposium on High-Performance Computer Architecture (HPCA)},
  pages={692--708},
  year={2022},
  organization={IEEE}
}

@article{tran2018spectral,
  title={Spectral signatures in backdoor attacks},
  author={Tran, Brandon and Li, Jerry and Madry, Aleksander},
  journal={Advances in neural information processing systems},
  volume={31},
  year={2018}
}

\end{document}